\documentstyle[aps,epsfig,amssymb,twocolumn]{revtex}
\begin{document}

\title{Neutral Evolution of Mutational Robustness}
\author{Erik van Nimwegen\thanks{Bioinformatics Group, University of
Utrecht, Padualaan 8, NL-3584-CH Utrecht, The Netherlands.}
and James P. Crutchfield}
\address{Santa Fe Institute, 1399 Hyde Park Road, Santa Fe, New Mexico 87501\\
\{erik,chaos\}@santafe.edu}
\author{Martijn Huynen$^*$}
\address{Biocomputing Group, EMBL, Meyerhofstrasse 1, 69012 Heidelberg, Germany\\
Martijn.Huynen@EMBL-Heidelberg.de}
\date{\today}
\maketitle

\begin{abstract}
  We introduce and analyze a general model of a population evolving
  over a network of selectively neutral genotypes. We show that the
  population's limit distribution on the neutral network is solely
  determined by the network topology and given by the principal
  eigenvector of the network's adjacency matrix. Moreover, the average
  number of neutral mutant neighbors per individual is given by the
  matrix spectral radius. This quantifies the extent to which
  populations evolve mutational robustness: the insensitivity of the
  phenotype to mutations. Since the average
  neutrality is independent of evolutionary parameters---such as,
  mutation rate, population size, and selective advantage---one can
  infer global statistics of neutral network topology using simple
  population data available from {\it in vitro} or {\it in vivo}
  evolution. Populations evolving on neutral networks of RNA secondary
  structures show excellent agreement with our theoretical
  predictions.
\end{abstract}

\section{Introduction}

Kimura's contention that a majority of genotypic change in evolution
is selectively neutral \cite{Kimurabook} has gained renewed attention
with the recent analysis of evolutionary optimization methods
\cite{Crutchfield&Mitchell95a,Nimw97a} and the discovery of {\it
  neutral networks} in genotype-phenotype models for RNA secondary
structure \cite{Font93a,GrunerEtal96c,Schuster&FontanaEtal94} and
protein structure \cite{BabajideEtal97,BastollaEtal98}. It was found
that collections of mutually neutral genotypes, which are connected
via single mutational steps, form extended networks that permeate
large regions of genotype space. Intuitively, a large degeneracy in
genotype-phenotype maps, when combined with the high connectivity
of (high-dimensional) genotype spaces, readily leads to such extended
neutral networks. This intuition is now supported by recent
theoretical results
\cite{Barnett97,Gavrilets&Gravner97,Reidys&Fraser95,Reidys&Stadler&Schuster97}.

In {\it in vitro} evolution of ribozymes, mutations responsible for an
increase in fitness are only a small minority of the total number of
accepted mutations \cite{WrightJoyce:97}. This indicates that, even in
adaptive evolution, the majority of point mutations is neutral. The
fact that only a minority of loci is conserved in sequences evolved
from a single ancestor similarly indicates a high degeneracy in
ribozymal genotype-phenotype maps \cite{LandweberProkovskaya}.
Neutrality is also implicated in experiments where RNA sequences
evolve a given structure starting from a range of different initial
genotypes \cite{EklandBartel}. More generally, neutrality in RNA and
protein genotype-phenotype maps is indicated by the observation that
their structures are much better conserved during evolution than their
sequences \cite{Gutell:93,Huynen:98}.

Given the presence of neutral networks that preserve structure or
function in sequence space, one asks, How does an evolving population
distribute itself over a neutral network? Can we detect and analyze
structural properties of neutral networks from data on biological or
{\it in vitro} populations? To what extent does a population evolve
toward highly connected parts of the network, resulting in sequences
that are relatively insensitive to mutations? Such
{\em mutational robustness} has been observed in biological RNA
structures \cite{Huynen:93,Wagner&Stadler99} and in simulations of the
evolution of RNA secondary structure \cite{Huynen:94}. However, an
analytical understanding of the phenomenon, the underlying mechanisms,
and their dependence on evolutionary parameters---such as, mutation
rate, population size, selection advantage, and the topology of the
neutral network---has up to now not been available.

Here we develop a dynamical model for the evolution of populations on
neutral networks and show analytically that, for biologically relevant
population sizes and mutation rates, a population's distribution over
a neutral network is determined solely by the network's topology.
Consequently, one can infer important structural information about
neutral networks from data on evolving populations, even without
specific knowledge of the evolutionary parameters. Simulations of the
evolution of a population of RNA sequences, evolving on a neutral
network defined with respect to secondary structure, confirm our
theoretical predictions and illustrate their application to inferring
network topology.

\section{Modeling Neutrality}

We assume that genotype space contains a neutral network of high, but
equal fitness genotypes on which the majority of a population is
concentrated and that the neighboring parts of genotype space consist
of genotypes with markedly lower fitness. The genotype space consists
of all sequences of length $L$ over a finite alphabet ${\cal A}$ of
$A$ symbols. The neutral network on which the population moves can be
most naturally regarded as a graph $G$ embedded in this genotype
space. The vertex set of $G$ consists of all genotypes that are on the
neutral network; denote its size by $|G|$. Two vertices are connected
by an edge if and only if they differ by a single point mutation.

We will investigate the dynamics of a population evolving on this
neutral network and analyze the dependence of several population
statistics on the topology of the graph $G$. With these results, we
will then show how measuring various population statistics enables one
to infer $G$'s structural properties.

For the evolutionary process, we assume a discrete-generation
selection-mutation dynamics with constant population size $M$.
Individuals on the neutral network $G$ have a fitness $\sigma$.
Individuals outside the neutral network have fitnesses that are
considerably smaller than $\sigma$. Under the approximations we use,
the exact fitness values for genotypes off $G$ turn out to be
immaterial. Each generation, $M$ individuals are selected with
replacement and with probability proportional to fitness and then
mutated with probability $\mu$. These individuals form the next
generation.

This dynamical system is a discrete-time version of Eigen's molecular
evolution model \cite{Eigen71}. Our analysis can be directly
translated to the continuous-time equations for the Eigen model. The
results remain essentially unchanged.

Although our analysis can be extended to more complicated mutation
schemes, we will assume that only single point mutations can occur at
each reproduction of an individual. With probability $\mu$ one of the
$L$ symbols is chosen with uniform probability and is mutated to one
of the $A-1$ other symbols. Thus, under a mutation, a genotype $s$
moves with uniform probability to one of the $L(A-1)$ neighboring
points in genotype space.

\subsection{Infinite-Population Solution}

The first step is to solve for the asymptotic distribution of the
population over the neutral network $G$ in the limit of very large
population sizes. 

Once the (infinite) population has come to equilibrium, there will be
a constant proportion $P$ of the population located on the network
$G$ and a constant average fitness $\langle f \rangle$ in the
population.  Under selection the proportion of individuals on the
neutral network increases from $P$ to $\sigma P / \langle f \rangle$.
Under mutation a proportion $\langle \nu \rangle$ of these individuals
remains on
the network, while a proportion $1-\langle \nu \rangle$ falls off the
neutral network to lower fitness. At the same time, a proportion $Q$ of
individuals located outside $G$ mutate {\em onto} the network so that
an equal proportion $P$ ends up on $G$ in the next generation. Thus, at
equilibrium, we have a balance equation:
\begin{equation}
\label{equi_condition}
P = \frac{\sigma}{\langle f \rangle} \langle \nu \rangle P + Q.
\end{equation}

In general, the contribution of $Q$ to $P$ is negligible. As mentioned
above, we assume that the fitness $\sigma$ of the network genotypes is
substantially larger than the fitnesses of those off the neutral
network and that the mutation rate is small enough so that the bulk
of the population is located on the neutral network. Moreover, since
their fitnesses are smaller than the average fitness $\langle f
\rangle$, only a fraction of the individuals off the network $G$
produces offspring for the next generation. Of this fraction, only a
small fraction mutates {\em onto} the neutral network $G$. Therefore,
we neglect the term $Q$ in Eq. (\ref{equi_condition}) and obtain:
\begin{equation}
\label{mut_sec_balance}
\frac{\sigma}{\langle f \rangle} \langle \nu \rangle = 1.
\end{equation}
This expresses the balance between selection expanding the
population on the network and deleterious mutations reducing it by
moving individuals off.

Under mutation an individual located at genotype $s$ of $G$ with vertex
degree $d_s$ (the number of neutral mutant neighbors) has
probability
\begin{equation}
\label{node_neutrality}
\nu_s = 1-\mu \left( 1 - \frac{d_s}{(A-1) L}\right)
\end{equation}
to remain on the neutral network $G$. If asymptotically a fraction
$P_s$ of the population is located at genotype $s$, then $\langle \nu
\rangle$ is simply the average of $\nu_s$ over the asymptotic
distribution on the network: $\langle \nu \rangle = \sum_{s \in G}
\nu_s P_s/P$. As Eq.  (\ref{node_neutrality}) shows, the average
$\langle \nu \rangle$ is simply related to the {\em population
  neutrality} $\langle d \rangle = \sum_{s \in G} d_s P_s/P$.
Moreover, using Eq. (\ref{mut_sec_balance}) we can directly relate the
population neutrality $\langle d \rangle$ to the average fitness
$\langle f \rangle$:
\begin{equation}
\label{av_degree_expression}
\langle d \rangle = L (A-1) \left[ 1 -\frac{\sigma - \langle f
    \rangle}{\mu \sigma} \right].
\end{equation}

Despite our not specifying the details of $G$'s topology, nor giving
the fitness values of the genotypes lying off the neutral network, one
can relate the population neutrality $\langle d \rangle$ of
the individuals on the neutral network directly to the average fitness
$\langle f \rangle$ in the population. It may seem surprising that
this is possible at all. Since the population consists partly of
sequences off the neutral network, one expects that the
average fitness is determined in part by the fitnesses of these
sequences. However, under the assumption that {\em back mutations}
from low-fitness genotypes off the neutral network onto $G$ are
negligible, the fitnesses of sequences outside $G$ only influence
the total proportion $P$ of individuals on the network, but not the
average fitness in the population.

Equation (\ref{av_degree_expression}) shows that the population
neutrality $\langle d \rangle$ can be inferred from the average fitness
and other parameters---such as, mutation rate. However, as we will now
show, the population neutrality $\langle d \rangle$ can also be
obtained independently, from knowledge of the topology of $G$ alone.

The asymptotic equilibrium proportions $\{ P_s \}$ of the population
at network nodes $s$ are the solutions of the simultaneous equations:
\begin{equation}
P_s = (1-\mu) \frac{\sigma}{\langle f \rangle} P_s +
\frac{\mu}{L(A-1)} \sum_{t \in [ s ]_G} \frac{\sigma}{\langle f
  \rangle} P_{t},
\label{red_eq_mot}
\end{equation}
where $[s]_G$ is the set of neighbors of $s$ that are also on the
network $G$. Using Eq. (\ref{av_degree_expression}), Eq.
(\ref{red_eq_mot}) can be rewritten as:
\begin{equation}
\label{formal_eigen_eq}
\langle d \rangle P_s = \left({\bf G} \cdot \vec{P} \right)_s,
\end{equation}
where ${\bf G}$ is the adjacency matrix of the graph $G$: 
\begin{equation}
{\bf G}_{st} = 
  \left\{
  \begin{array}{ll}
  1 & t \in [s]_G,\\
  0 & \text{otherwise}. 
  \end{array}
  \right.
\end{equation}
Since ${\bf G}$ is nonnegative and the neutral
network $G$ is connected, the adjacency matrix is irreducible.
Therefore, the theorems of Frobenius-Perron for nonnegative
irreducible matrices apply\cite{Gantmacherbook}. These imply that the
proportions $P_s$ of the limit distribution on the network are given
by the principal eigenvector of the graph adjacency matrix ${\bf G}$.
Moreover, the population neutrality is equal to ${\bf G}$'s spectral
radius $\rho$: $\langle d \rangle = \rho$. In this way, one concludes
that asymptotically the population neutrality $\langle d \rangle$
is independent of evolutionary parameters ($\mu$,
$L$, $\sigma$) and of the fitness values of the genotypes off the
neutral network. It is a function {\em only} of the neutral network
topology as determined by the adjacency matrix ${\bf G}$.

This fortunate circumstance allows us to consider several practical
consequences. Note that knowledge of $\mu$, $\sigma$, and $\langle f
\rangle$ allows one to infer a dominant feature of $G$'s topology,
namely, the spectral radius $\rho$ of its adjacency matrix. In
{\it in vitro} evolution experiments in which biomolecules are evolved
(say) to bind a particular ligand \cite{Turk:97}, by measuring the
proportion $\langle \nu \rangle$ of functional molecules that remain
functional after mutation, we can now infer the spectral radius $\rho$
of their neutral network. In other situations, such as in the bacterial
evolution experiments of Ref. \cite{Elena&Cooper&Lenski96}, it might
be more natural to measure the average fitness $\langle f \rangle$ of an
evolving population and then use Eq. (\ref{av_degree_expression}) to
infer the population neutrality $\langle d \rangle$ of viable genotypes
in sequence space.

\subsection{Blind and Myopic Random Neutral Walks}

In the foregoing we solved for the asymptotic average neutrality
$\langle d \rangle$ of an infinite population under selection
and mutation dynamics and showed that it was uniquely determined by
the topology of the neutral network $G$. To put this result in
perspective, we now compare the population neutrality $\langle d
\rangle$ with the effective neutralities observed under two different
kinds of random walk over $G$. The results illustrate informative
extremes of how network topology determines the population dynamics
on neutral networks.

The first kind of random walk that we consider is generally referred
to as a {\em blind ant} random walk. An ant starts out on a randomly
chosen node of $G$. Each time step it chooses one of its
$L(A-1)$ neighbors at random. If the chosen neighbor is on $G$, the ant
steps to this node, otherwise it remains at the current node for another
time step. It is easy to show that this random walk asymptotically
spends equal amounts of time at all of $G$'s nodes \cite{HughesbookII}.
Therefore, the {\it network neutrality} $\bar{d}$ of the nodes visited
under this type of random walk is simply given by:
\begin{equation}
\label{blind_ant_d}
\bar{d} = \sum_{s \in G} \frac{d_s}{|G|}.
\label{AvNetworkNeutrality}
\end{equation}

It is instructive to compare this with the effective neutrality
observed under another random walk, called the {\em myopic ant}. An
ant again starts at a random node $s \in G$. Each time step, the ant
determines the set $[s]_G$ of network neighbors of $s$ and then steps
to one at random. Under this random walk, the asymptotic proportion
$P_s$ of time spent at node $s$ is proportional to the node degree
$d_s$ \cite{HughesbookII}. It turns out that the {\it myopic neutrality}
$\widehat{d}$ seen by this ant can be expressed in terms of the mean
$\bar{d}$ and variance ${\rm Var}(d)$ of node degrees over $G$:
\begin{equation}
\widehat{d} = \bar{d} + \frac{{\rm Var}(d)}{\bar{d}}.
\label{alt_myopic_ant_d}
\end{equation}
The network and myopic neutralities, $\bar{d}$ and $\widehat{d}$, are
thus directly given in terms of {\em local} statistics of the
distribution of vertex degrees, while the population neutrality
$\langle d \rangle$ is given by $\rho$, the spectral radius of $G$'s
adjacency matrix. The latter is an essentially {\em global} property
of $G$.

\section{Mutational Robustness}

With these cases in mind, we now consider how different network
topologies are reflected by these neutralities. In prototype models of
populations evolving on neutral networks, the networks are often
assumed to be or are approximated as regular graphs
\cite{Forst&Reidys&Weber95,ReidysPhd,Reidys&Stadler&Schuster97,Nimw97a,Nimw97b}.
If the graph $G$ is, in fact, regular, each node has the same degree $d$
and, obviously, one has $\langle d \rangle = \bar{d} = \widehat{d} = d$.

In more realistic neutral networks, one expects $G$'s neutralities to
vary over the network. When this occurs, the population neutrality is
typically larger than the network neutrality: $\langle d \rangle =
\rho > \bar{d}$. This difference quantifies precisely the extent to
which a population seeks out the most connected areas of the neutral
network. Thus, a population will evolve a {\em mutational robustness}
that is larger than if the population were to spread uniformly over
the neutral network. Additionally, the mutational robustness tends to
increase during the transient phase in which the population relaxes
towards the its asymptotic distribution.

Assume, for instance, that initially the population is located
entirely off the neutral network $G$ at lower fitness sequences. At
some time, a genotype $s \in G$ is discovered by the population. To a
rough approximation, one can assume that the probability of a genotype
$s$ being discovered first is proportional to the number of neighbors,
$L (A-1) - d_s$, that $s$ has {\em off} the neutral network.
Therefore, the population neutrality $\langle d_0 \rangle$ when the
population first enters the neutral network $G$ is approximately given
by:
\begin{equation}
\langle d_0 \rangle = \bar{d} - \frac{{\rm Var}(d)}{L(A-1) - \bar{d}}.
\end{equation} 
Therefore, we define the {\em excess robustness} $r$ to be the
relative increase in neutrality between the initial neutrality and
(asymptotic) population neutrality:
\begin{equation}
\label{robust_def}
r \equiv \frac{\langle d \rangle-\langle d_0 \rangle}{\langle d_0 \rangle}.
\end{equation}
For networks that are sparse, i.e. $\bar{d} \ll L(A-1)$, this is well
approximated by $r \approx (\langle d \rangle-\bar{d})/\bar{d}$. Note
that, while $r$ is defined in terms population statistics, the
preceding results have shown that this robustness is only a function
of $G$'s topology and should thus be considered a property of the
network.

\section{Finite-Population Effects}

Our analysis of the population distribution on the neutral network $G$
assumed an infinite population. For finite populations, it is well known
that sampling fluctuations converge a population and this raises a
question: To what extent does the asymptotic distribution $P_s$ still
describe the distribution over the network for small populations? As a
finite population diffuses over a neutral network
\cite{Huynen&Stadler&Fontana}, one might hope that the time average of
the distribution over $G$ is still given by $P_s$. Indeed, the
simulation results shown below indicate that for moderately large
population sizes, this seems to be the case. However, a simple argument
shows that this cannot be true for arbitrarily small populations.

Assume that the population size $M$ is so small that the product of
mutation rate and population size is much smaller than $1$; i.e. $M
\mu \ll 1$. In this limit the population will, at any point in time,
be completely converged onto a single genotype $s$ on the neutral
network $G$. With probability $M \mu$ a single mutant will be produced
at each generation. This mutant is equally likely to be one of the
$L(A-1)$ neighbors of $s$. If this mutant is not on $G$, it will
quickly disappear due to selection. However, if the mutant is on the
neutral network, there is a probability $1/M$ that it will take over
the population. When this happens, the population will effectively
have taken a random-walk step on the network, of exactly the kind
followed by the blind ant. Therefore, for $M \mu \ll 1$, the
population neutrality will be equal to the network neutrality:
$\langle d \rangle = \bar{d}$. In this regime, $r \approx 0$ and
excess mutational robustness will not emerge through evolution.

The extent to which the initial population neutrality approaches
$\langle d \rangle$ is determined by the extent to which evolution on
$G$ is dominated by sampling fluctuations. In neutral evolution,
population convergence is generally only a function of the product
$M \mu$ \cite{Derrida&Peliti,Kimura64,Wright31}. Thus, as the product
$M\mu$ ranges from values much smaller than $1$ to values much larger
than $1$, we predict that the population neutrality $\langle d \rangle$
shifts from the network neutrality $\bar{d}$ to the infinite-population
neutrality, given by $\bf G$'s spectral radius $\rho$.

\section{RNA Evolution on Structurally Neutral Networks} 

The evolution of RNA molecules in a simulated flow reactor provides an
excellent arena in which to test the theoretical predictions of
evolved mutational robustness. The replication rates (fitnesses) were
chosen to be a function only of the secondary structures of the RNA
molecules.  The secondary structure of RNA is an essential aspect of
its phenotype, as documented by its conservation in evolution
\cite{Gutell:93} and the convergent {\it in vitro} evolution toward a
similar secondary structure when selecting for a specific function
\cite{EklandBartel}. RNA secondary structure prediction based on free
energy minimization is a standard tool in experimental biology and has
been shown to be reliable, especially when the minimum free energy
structure is thermodynamically well defined \cite{Huynen:97}.  RNA
secondary structures were determined with the Vienna Package
\cite{Hofacker:94a}, which uses the free energies from
\cite{Walter:94}. Free energies of dangling ends were set to $0$.

\begin{figure}[htbp]
\centerline{\epsfig{file=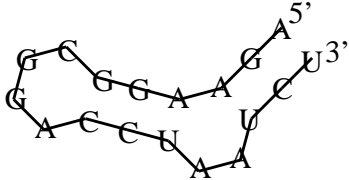,height=1.0in}}
\caption{The target RNA secondary structure.}
\label{structure_fig}
\end{figure}

The neutral network $G$ on which the population evolves consists of
all RNA molecules of length $L=18$ that fold into a particular
{\em target structure}. A target structure (Fig. \ref{structure_fig})
was selected that contains sufficient variation in its neutrality
to test the theory, yet is not so large as to preclude an exhaustive
analysis of its neutral network topology.

Using only single point mutations per replication, purine-pyrimidine
base pairs \{G-C, G-U, A-U\} can mutate into each other, but not into
pyrimidine-purine \{C-G, U-G, U-A\} base pairs. The target structure
contains $6$ base pairs which can each be taken from one or the other
of these two sets. Thus, the approximately $2 \times 10^8$ sequences
that are consistent with the target's base pairs separate into
$2^6=64$ disjoint sets. Of these, we analyzed the set in which all
base pairs were of the purine-pyrimidine type and found that it
contained two neutral networks of $51,028$ and $5,169$ sequences that
fold into the target structure. Simulations were performed on the
largest of the two. The exhaustive enumeration of this network showed
that it had a network neutrality of $\bar{d} = 12.0$ with standard
deviation $\sqrt{Var(d)} \approx 3.4$, a maximum neutrality of $d_s =
24$, and a minimum of $d_s = 1$. The spectral radius of the network's
$51028 \times 51028$ adjacency matrix was $\rho \approx 15.7$.  The
theory predicts that, when $M\mu \gg 1$, the population neutrality
should converge to this value.

The simulated flow reactor contained a population of replicating and
mutating RNA sequences \cite{Eigen71,fontana:89}. The replication rate
of a molecule depends on whether its calculated minimum free energy
structure equals that of the target: Sequences that fold into the
target structure replicate on average once per time unit, while all
other sequences replicate once per $10^4$ time units on average.
During replication the progeny of a sequence has probability $\mu$ of
a single point mutation. Selection pressure in the flow reactor is
induced by an adaptive dilution flow that keeps the total RNA
population fluctuating around a constant capacity $M$.

\begin{figure}[htbp]
\centerline{\epsfig{file=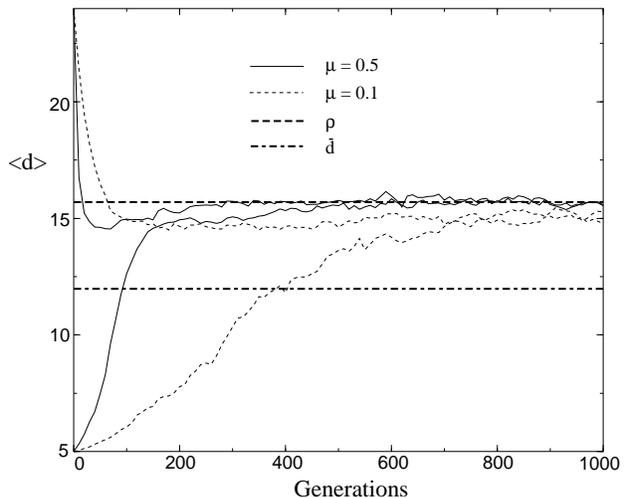,width=3.25in}}
\caption{The evolution of RNA mutational robustness: convergence of the
  population's average neutrality to the theoretical value,
  $\langle d \rangle = \rho \approx 15.7$, predicted by
  $\bf G$'s spectral radius (upper dashed line). The network's average
  neutrality $\bar{d}$ is the lower dashed line. Simulations used a
  population size of $M = 10^4$ and mutation rates of $\mu=0.5$ and
  $\mu=0.1$ per sequence. They were started at sequences with either
  a relatively large number of neutral neighbors ($d_s = 24$)
  (upper curves for each mutation rate) or a small number
  ($d_s = 5$) (lower curves).
  }
\label{two_runs}
\end{figure}

Evolution was seeded from various starting sequences with either a
relatively high or a relatively low neutrality. Independent of the
starting point, the population neutrality converges to the predicted
value, as shown in Fig. \ref{two_runs}.

\begin{figure}[htbp]
\centerline{\epsfig{file=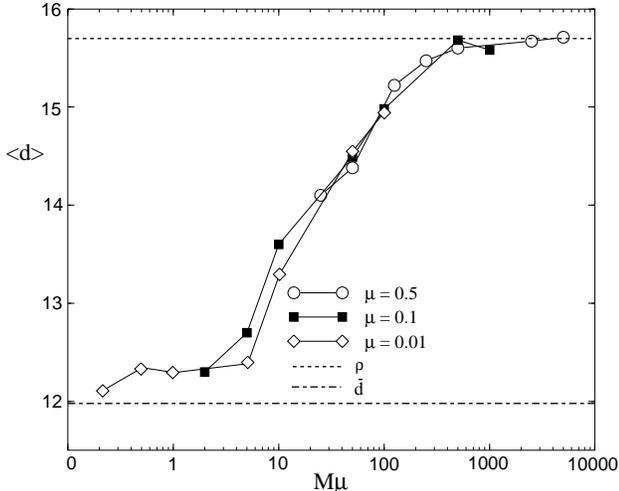,width=3.25in}}
\caption{Dependence of the average neutrality in the population on
  mutation rate $\mu$ and population size $M$. Simulations used three
  mutation rates, $\mu \in \{0.5, 0.1, 0.01\}$, and a range of
  population sizes, $M \in \{10000,5000,1000,500,250,100,50,20\}$.
  The results show that the evolved neutrality in the population
  depends on the product $M \mu$ of population size and mutation rate.
  Neutrality increases with increasing $M \mu$ and saturates when
  $M \mu > 500$. When $M \mu < 1$ population neutrality approximates
  $G$'s average neutrality $\bar{d} \approx 12.0$. When $M \mu > 500$
  population neutrality converges to the spectral radius of the
  network's adjacency matrix, $\rho \approx 15.7$.}
\label{Mmu_dependence}
\end{figure}

Subsequently, we tested the finite-population effects on the
population's average neutrality at several different mutation rates.
Figure \ref{Mmu_dependence} shows the dependence of the asymptotic
average population neutrality on population size $M$ and mutation rate
$\mu$. As expected, the population neutrality depends only on the
product $M\mu$ of population size and mutation rate. For small $M\mu$
the population neutrality increases with increasing $M\mu$, until
$M\mu \approx 500$ where it saturates at the predicted value of
$\langle d \rangle \approx 15.7$.  Since small populations do not form a
stationary distribution over the neutral net, but diffuse over it
\cite{Huynen&Stadler&Fontana}, the average population neutrality at
each generation may fluctuate considerably for small populations.
Theoretically, sampling fluctuations in the proportions of individuals
at different nodes of the network scale inversely proportional to the
square root of the population size. We therefore expect the fluctuations
in population neutrality to scale as the inverse square root
of the population size as well. This was indeed observed in our
simulations.

Finally, the fact that $r \approx 0.31$ for this neutral network shows
that under selection and mutation, a population will evolve a
mutational robustness that is $31$ percent higher than if it were to
spread randomly over the network.

\section{Conclusions}

We have shown that, under neutral evolution, a population does not
move over a neutral network in an entirely random fashion, but tends
to concentrate at highly connected parts of the network, resulting
in phenotypes that are relatively robust against mutations. Moreover,
the average number of point mutations that leave the phenotype unaltered
is given by the spectral radius of the neutral network's adjacency
matrix. Thus, our theory provides an analytical foundation for the
intuitive notion that evolution selects genotypes that are
mutationally robust.

Perhaps surprisingly, the tendency to evolve toward highly connected
parts of the network is independent of evolutionary parameters---such
as, mutation rate, selection advantage, and population size (as long as
$M \mu \gg 1$)---and is solely determined by the network's topology.
One consequence is that one can infer properties of the neutral
network's topology from simple population statistics.

Simulations with neutral networks of RNA secondary structures confirm
the theoretical results and show that even for moderate population
sizes, the population neutrality converges to the infinite-population
prediction. Typical sizes of {\it in vitro} populations are such that
the data obtained from experiments are expected to accord with the
infinite-population results derived here. It seems possible then to
devise {\it in vitro} experiments that, using the results outlined
above, would allow one to obtain information about the topological
structure of neutral networks of biomolecules with similar
functionality.

We will present the extension of our theory to cases with
multiple-mutation events per reproduction elsewhere. We will also
report on analytical results for a variety of network topologies that
we have studied.

Finally, here we focused only on the asymptotic distribution of the
population on the neutral network. But how did the population attain
this equilibrium? The transient relaxation dynamics,
such as that shown in Fig. \ref{two_runs}, can be analyzed in terms of
the nonprincipal eigenvectors and eigenvalues of the adjacency matrix
${\bf G}$. Since the topology of a graph is almost entirely determined
by the eigensystem of its adjacency matrix, one should in principle be
able to infer the complete structure of the neutral network from
accurate measurements of the transient population dynamics.

{\bf Acknowledgments} We thank the participants of the Santa Fe
Institute workshop on Evolutionary Dynamics for stimulating this work,
which was supported in part at SFI by NSF under grant IRI-9705830,
Sandia National Laboratory, and the Keck Foundation. M.H. gratefully
acknowledges support from a fellowship of the Royal Netherlands Academy
of Arts and Sciences.

\bibliography{epev} 
\bibliographystyle{plain}

\end{document}